\documentclass[
aip,
amsmath,amssymb,
reprint,
]{revtex4-1}

\usepackage{graphicx}
\usepackage{dcolumn}
\usepackage{bm}
\usepackage{booktabs}
\usepackage{empheq}
\usepackage{physics}
\usepackage{siunitx}
\usepackage[utf8]{inputenc}
\usepackage[T1]{fontenc}
\usepackage{mathptmx,etoolbox,textcomp,gensymb,upgreek}

\newcommand{\xmin}{\tilde{x}_{\text{min}}}

\newcommand{\um}{\upmu\text{m}}

\newcommand{\kb}{k_\mathrm{B}}

\makeatletter
\def\@email#1#2{%
\endgroup
\patchcmd{\titleblock@produce}
{\frontmatter@RRAPformat}
{\frontmatter@RRAPformat{\produce@RRAP{*#1\href{mailto:#2}{#2}}}\frontmatter@RRAPformat}
{}{}
}%
\makeatother

\begin{document}

\preprint{AIP/123-QED}

\title{Photon-Number Discrimination from Increasing Time-of-Arrival
Separation in Wide Superconducting Microstrip Single-Photon Detectors}

\author{Hsin-Yeh Wu}
\email{wuhsinyeh@ntu.edu.tw}
\affiliation{Department of Physics, National Taiwan University, Taipei 10617, Taiwan}
\affiliation{Leung Center for Cosmology and Particle Astrophysics, National Taiwan University, Taipei 10617, Taiwan}
\author{Shu-Xiao Liu}
\affiliation{Department of Physics, National Taiwan University, Taipei 10617, Taiwan}
\affiliation{Research Center for Applied Sciences, Academia Sinica, Taipei 11529, Taiwan}
\author{Boris Tuchming}
\affiliation{CEA-IRFU, Paris-Saclay University 91190 Gif sur Yvette cedex, France}
\author{Xavier-François Navick}
\affiliation{CEA-IRFU, Paris-Saclay University 91190 Gif sur Yvette cedex, France}
\author{Pu-Kai Wang}
\affiliation{Institute of Physics, Academia Sinica, Taipei 11529, Taiwan}
\author{Yi-Ren Wu}
\affiliation{Department of Physics, National Taiwan University, Taipei 10617, Taiwan}
\author{Jia-Wern Chen}
\affiliation{Research Center for Applied Sciences, Academia Sinica, Taipei 11529, Taiwan}
\author{Marc Besançon}
\affiliation{CEA-IRFU, Paris-Saclay University 91190 Gif sur Yvette cedex, France}
\author{Jean-François Glicenstein}
\affiliation{CEA-IRFU, Paris-Saclay University 91190 Gif sur Yvette cedex, France}
\author{Pisin Chen}
\affiliation{Department of Physics, National Taiwan University, Taipei 10617, Taiwan}
\affiliation{Leung Center for Cosmology and Particle Astrophysics, National Taiwan University, Taipei 10617, Taiwan}
\author{Yu-Jung Lu}
\affiliation{Department of Physics, National Taiwan University, Taipei 10617, Taiwan}
\affiliation{Research Center for Applied Sciences, Academia Sinica, Taipei 11529, Taiwan}
\author{Stathes Paganis}
\email{paganis@phys.ntu.edu.tw}
\affiliation{Department of Physics, National Taiwan University, Taipei 10617, Taiwan}
\affiliation{Leung Center for Cosmology and Particle Astrophysics, National Taiwan University, Taipei 10617, Taiwan}

\date{\today}

\begin{abstract}
Measurements of photon-number-dependent time-of-arrival (TOA)
distributions in wide NbN superconducting microstrip single-photon
detectors (SMSPDs) are presented. Two \SI{10}{\um}-long devices with
widths of \SI{3.2}{\um} and \SI{6.4}{\um} are characterized under
pulsed \SI{515}{nm} excitation. The TOA histograms are described by
sums of exponentially modified Gaussian (EMG) components, with the
wider device exhibiting components assigned to photon numbers up to
\(n=8\). Adjacent TOA peaks become more distinguishable relative to
their widths at higher photon number, in contrast to the compression
of neighboring photon-number timing shifts observed in conventional
narrow-wire SNSPDs. For the components assigned to \(n=7\) and
\(n=8\), the FWHM-normalized peak separation is \(2.15\),
and the fitted EMG distributions yield an equal-prior pairwise
TOA-assignment fidelity of \(99.6\%\). The EMG exponential-tail 
parameter \(\tau_n\) decreases strongly with photon number, while 
the Gaussian-location parameter \(\mu_n\) continues to shift between
neighboring components. A phenomenological
model motivated by absorption-site-dependent vortex processes and
collective current redistribution captures these timing trends.
These results support wide SMSPDs as a platform for TOA-based
photon-number discrimination at high photon number, relevant to
multiphoton-state characterization and photonic quantum information
processing. The same geometry also provides a setting to examine how the
spatial distribution of photon absorption affects vortex-assisted
detection.
\end{abstract}

\maketitle

\section{Introduction}

Superconducting nanowire single-photon detectors (SNSPDs) combine high
detection efficiency, low timing jitter, and low dark-count rates, and
have become essential tools in quantum photonics, quantum communication,
and precision optical measurements
~\cite{eszadeh2021,pirandola2020,reddy2020,chang2021}. In many
applications, detecting the presence of light is not sufficient; one
also needs to discriminate among optical events associated with
different photon numbers. Photon-number discrimination is important for
source and state characterization, detector calibration, and photonic
quantum-information applications~\cite{xie2024,los2024}. Higher
photon-number capability is particularly valuable for characterizing
multiphoton optical states and measuring high-order photon statistics
~\cite{cheng2023}.

Photon-number-sensitive superconducting detectors have been realized
with multiplexed architectures, which extend the accessible
photon-number range at the cost of additional detector elements and
readout complexity. Intrinsic single-element approaches have also been
demonstrated in conventional narrow-wire SNSPDs, where multiple absorbed
photons modify the pulse amplitude or timing response. For example,
impedance-engineered readout enhances pulse-amplitude sensitivity and
enables few-photon discrimination in a single nanowire
~\cite{cahall2017,zhu2020}, while photon-number-dependent turn-on
dynamics have also been studied in conventional SNSPDs
~\cite{nicolich2019}. A complementary single-device approach uses
dual-mode operation: an SNSPD can operate in either a Geiger counting
mode or an energy-sensitive calorimetric mode
~\cite{wu2025dualmode}.

In conventional narrow-wire SNSPDs, the response can occur earlier for
larger deposited energy, for instance, when resistive-domain formation is
accelerated. This produces photon-number-dependent TOA components that
can be used for timing-based discrimination. At the
microscopic level, photon detection in these devices is commonly
described in terms of photon-induced local suppression of
superconductivity, current redistribution, and the formation of a
resistive response. In vortex-assisted descriptions, the local
perturbation lowers the barrier for vortex entry from a strip edge, and a
vortex driven across the strip can trigger the dissipative transition
associated with the electrical readout pulse~\cite{Bulaevskii_2011,
bulaevskii2012,engel2015,vodolazov2015,korneeva2018}.
Despite this photon-number-dependent timing response, the spacing
between neighboring TOA peaks in narrow-wire SNSPDs can decrease at
higher photon number, making timing-based discrimination more
challenging~\cite{sidorova2025}.

Wide, short SMSPDs provide a distinct geometry with a wide transverse
dimension and a limited length along the strip. Characteristic
hotspot-interaction length scales reported for NbN are of order
\SI{25}{nm}~\cite{renema2017}, far smaller than the strip
widths studied here. Photon absorption may therefore initially perturb
the superconducting state locally rather than immediately produce a
belt-like excitation across the full strip width. In longer strips or
meandered devices, photon absorptions can be distributed over separated
longitudinal portions of the strip. The limited length of the present
strips restricts the range of absorption-site separations along the
strip, while their width retains a substantial transverse spatial degree
of freedom. Transverse absorption position is therefore a natural
variable for interpreting the photon-number-dependent TOA response.

In this work, we study two \SI{10}{\um}-long NbN SMSPDs with widths of
\SI{3.2}{\um} and \SI{6.4}{\um} under pulsed illumination at
\SI{515}{nm}. Their TOA histograms are fitted with sums of exponentially 
modified Gaussian (EMG) components assigned to photon numbers, 
with resolved components extending through \(n=5\) in the narrower device 
and through \(n=8\) in the wider device. The adjacent fitted TOA
components become more separable at high photon number. For the
components assigned to \(n=7\) and \(n=8\) in the wider device, the
FWHM-normalized peak separation is
\(\mathcal{R}_{7,8}^{\mathrm{FWHM}}=2.15\), corresponding to an
equal-prior pairwise TOA assignment fidelity of
\(F_{7,8}^{\mathrm{pair}}=99.6\%\).

The distinct photon-number dependences of the EMG exponential-tail scale
and Gaussian-location parameter are captured by a phenomenological
two-contribution model motivated by edge-sensitive vortex processes and
the collective effect of multiple absorbed photons. The model provides a
physically motivated interpretation of the measured TOA trends, rather
than a unique microscopic decomposition of the fitted timing parameters.

The remainder of this paper is organized as follows. Section~II
describes the devices and measurements. Section~III introduces the EMG
decomposition and the phenomenological timing model. Section~IV presents
the extracted TOA parameters and adjacent-component discrimination
metrics. Section~V discusses the physical interpretation, the influence
of finite beam width, and implications for future device optimization.

\begin{figure*}
\centering
\includegraphics[width=0.98\textwidth]{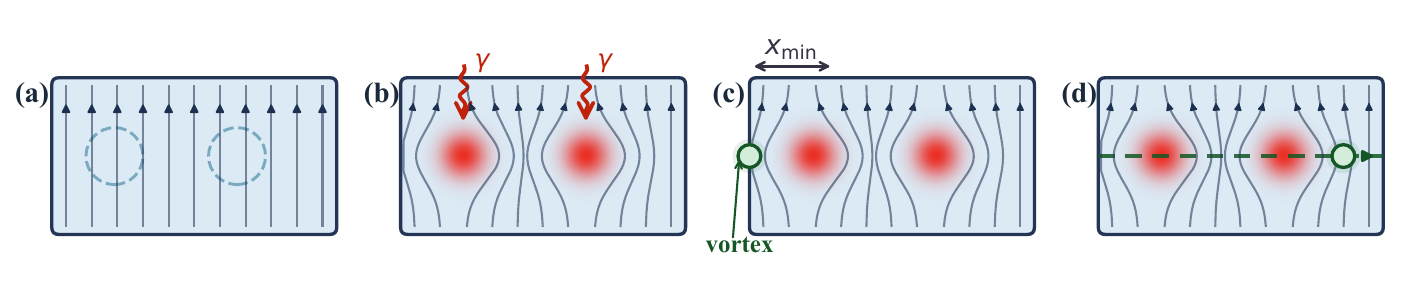}
\caption{
Schematic illustration of the edge-sensitive nucleation and
subsequent vortex-traversal processes considered in the timing
description.
(a) Bias current flows uniformly in the superconducting
microstrip before photon absorption. (b) Absorbed photons generate localized
hotspots and redistribute the current density. (c) A vortex nucleates
stochastically at the strip edge, with the relevant transverse coordinate
set by the edge-nearest hotspot, \(\xmin\). (d) The vortex then traverses
the strip, producing the dissipative event associated with the electrical
readout pulse.}
\label{fig:1}
\end{figure*}

\section{Devices and Methods}
\label{sec:methods}

The devices were fabricated from \SI{15}{nm}-thick NbN films deposited
on sapphire substrates by radio-frequency sputtering at \SI{120}{W},
with an Ar:N$_2$ flow ratio of 36:0.3~sccm, a chamber pressure of
\SI{0.9}{mTorr}, and a substrate temperature of \SI{800}{\celsius}.
The films had a superconducting critical temperature of approximately
\SI{9}{K}. The microstrips were patterned by laser direct writing using
KL-5310 photoresist, developed in AD-10 developer, and etched by
reactive-ion etching with a CF$_4$:O$_2$ flow ratio of 20:2~sccm and a
direct-current power of \SI{200}{W}. A \SI{5}{nm}-thick Al$_2$O$_3$
capping layer was subsequently deposited by atomic layer deposition.

Two devices fabricated from the same NbN film were characterized. Both
devices have a length of \SI{10}{\um}. Device~A has a width of
\SI{3.2}{\um}, whereas Device~B has a width of \SI{6.4}{\um}. Their
factor-of-two difference in width is used to examine the width
dependence of the photon-number-dependent TOA response and
adjacent-photon-number TOA discrimination.

The devices were mounted on the cold stage of a cryogenic system and
operated at \SI{6}{K}. Device~A had a switching current
\(I_\mathrm{sw}=\SI{90}{\uA}\), while Device~B had
\(I_\mathrm{sw}=\SI{220}{\uA}\); both devices were biased at
approximately \(0.80\,I_\mathrm{sw}\).

Pulsed excitation was supplied by a PicoQuant Prima three-color
gain-switched picosecond laser operated at \SI{515}{nm} in Narrow Pulse
mode, with a repetition rate of \SI{100}{kHz}. The specified pulse
width of the \SI{515}{nm} output in this mode is below
\SI{170}{ps} FWHM.  The laser was coupled through free-space optics 
and focused onto the active region using an attocube LT-APO/VIS/0.82
objective. The estimated Gaussian-equivalent beam waist
is $w_0\approx0.3~\mu\mathrm{m}$. The incident optical power was varied while
the optical alignment, device bias, and readout configuration were held
fixed.

The devices were biased through a bias tee
(Mini-Circuits ZFBT-4R2GW+). The output pulses were amplified with a
room-temperature low-noise amplifier (RFBay LNA-1030; bandwidth
20~MHz--1~GHz, gain 30~dB, noise figure 1.3~dB) and recorded using a
PicoHarp~300 time-correlated single-photon-counting module. The TOA was
measured relative to the laser synchronization signal. The PicoHarp~300
uses programmable constant-fraction discrimination on both the
synchronization and detector channels, reducing sensitivity of the
timing pickoff to pulse-amplitude variation relative to fixed
leading-edge discrimination. TOA histograms were recorded with a
\SI{4}{ps} time bin. This bin width is distinct from the overall timing
response, which also includes contributions from the laser
synchronization output, detector, amplification chain, CFD timing
pickoff, and TCSPC electronics.

\section{EMG Decomposition and Phenomenological Model}
\label{sec:model}

To describe the measured TOA histograms, we first use an exponentially
modified Gaussian (EMG) decomposition to extract the timing parameters
of the resolved TOA components. Following the analysis of
photon-number-dependent timing histograms in conventional meander SNSPDs
in Ref.~\cite{sidorova2025}, we use the EMG form to describe the observed
asymmetric TOA distributions. We then introduce a phenomenological
description of the photon-number dependence of the fitted EMG
parameters. Motivated by vortex-assisted detection and current
redistribution~\cite{engel2015,vodolazov2015}, 
this description is developed for the wide, short
microstrip geometry studied here. It provides a physical interpretation
of the timing trends rather than a microscopic derivation of the
detection process.

The measured TOA histogram is
represented as a finite mixture of normalized EMG components,
\begin{equation}
H(t)=C+\sum_{n=1}^{N_\mathrm{max}}
A_n\,P_\mathrm{EMG}(t;\mu_n,\sigma_n,\tau_n),
\label{eq:hist_model}
\end{equation}
where \(C\) is a constant background term and \(A_n\) is the integrated
area of the component assigned to photon number \(n\). The
corresponding normalized EMG density is
\(p_n(t)=P_\mathrm{EMG}(t;\mu_n,\sigma_n,\tau_n)\), with
\begin{equation}
\begin{aligned}
P_\mathrm{EMG}(t;\mu_n,\sigma_n,\tau_n)
&=
\frac{1}{2\tau_n}
\exp\!\left(
\frac{\sigma_n^2}{2\tau_n^2}
-\frac{t-\mu_n}{\tau_n}
\right) \\
&\quad\times
\operatorname{erfc}\!\left[
\frac{\sigma_n^2/\tau_n-(t-\mu_n)}
{\sqrt{2}\,\sigma_n}
\right].
\end{aligned}
\label{eq:EMG}
\end{equation}
Here, \(\mu_n\) is the Gaussian-location parameter, \(\sigma_n\) is
its standard deviation, and \(\tau_n\) is the exponential-tail scale.
The peak time \(t_n^\mathrm{pk}\) and full width at half maximum
\(\mathrm{FWHM}_n\) are evaluated numerically from the fitted EMG
component, with \(t_n^\mathrm{pk}\) defined by the maximum of \(p_n(t)\).
The EMG decomposition is used here as an empirical description of the
measured component shapes; the physical interpretation of the fitted
parameters is introduced below.

The fitted TOA components are assigned to photon number \(n\) using their
temporal ordering and the optical-power dependence of their integrated
areas \(A_n\). The normalized component areas are compared with a
Poisson model over the included range \(1\leq n\leq N_\mathrm{max}\),
which describes the photon-number statistics per pulse. The
fitted Poisson mean, \(\langle n\rangle_\mathrm{fit}\),
characterizes the component populations at each optical power. Details
of the Poisson fit are given in the Supplementary Material.

To interpret the photon-number dependence of \(\tau_n\) and \(\mu_n\),
we use a two-part description of the measured delay,
\begin{equation}
\mathrm{TOA}
=
\mu_c+T_\mathrm{nuc}(\xmin)+T_\mathrm{trav}(n),
\label{eq:TOA_decomp}
\end{equation}
Here, \(\mu_c\) is a fixed timing offset of the measurement chain,
\(T_\mathrm{nuc}\) is an effective stochastic contribution associated
with vortex nucleation near a strip edge, and \(T_\mathrm{trav}\) is an
effective contribution associated with subsequent vortex traversal
across the strip. Vortex-assisted detection, in which local photon
absorption and current redistribution reduce the barrier for vortex
entry, provides the physical motivation for this decomposition
~\cite{Bulaevskii_2011,bulaevskii2012,engel2015,vodolazov2015,korneeva2018}.
Figure~\ref{fig:1} schematically summarizes these processes. In the
phenomenological mapping used below, the fitted tail scale \(\tau_n\)
parameterizes the nucleation-associated stochastic contribution,
whereas \(\mu_n-\mu_c\) parameterizes the traversal-associated location
contribution. These contributions are not independently resolved by the
TOA data.

For the nucleation contribution, we introduce an idealized transverse
description in which absorption positions are independent and uniformly
distributed across the strip width. The folded nearest-edge coordinate
\(\tilde{x}\in[0,1]\) is defined by \(\tilde{x}=0\) at either edge and
\(\tilde{x}=1\) at the strip center. This construction is motivated by
the reported transverse dependence of SNSPD response, including enhanced
edge response in NbN nanowires~\cite{Renema2015,engel2015}. For \(n\)
absorbed photons, the relevant coordinate is
\(\xmin\equiv\min(\tilde{x}_1,\ldots,\tilde{x}_n)\). For independent
absorption positions uniformly distributed across the strip width,
\(\xmin\) has density \(n(1-\tilde{x})^{n-1}\) on \([0,1]\), with
\(\mathbb{E}[\xmin]=1/(n+1)\).

\begin{figure*}[t]
\centering
\includegraphics[width=1\textwidth]{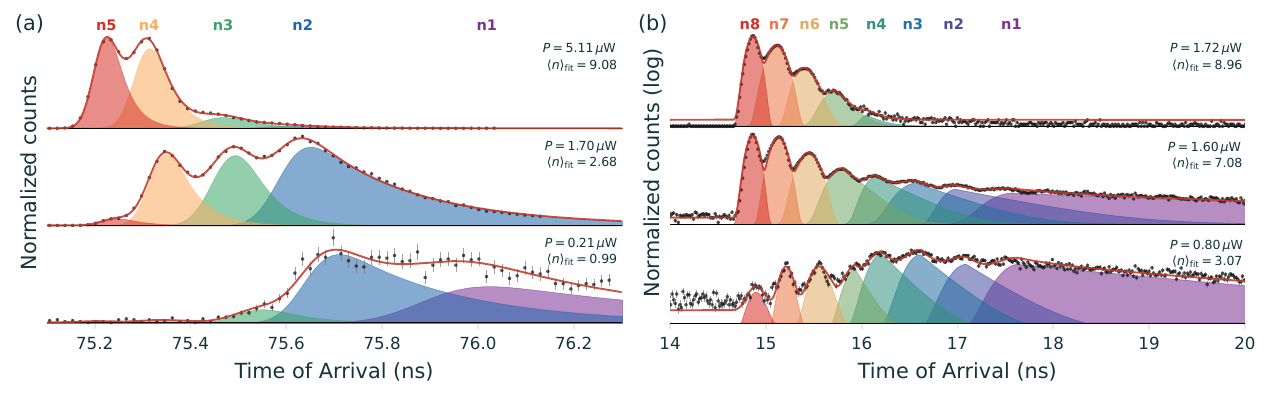}
\caption{Representative stacked TOA histograms for (a) Device~A and
(b) Device~B, with EMG decomposition overlays. Three representative
incident optical powers are shown in each panel, increasing from bottom
to top. The annotations give the incident optical power \(P\) and the
fitted Poisson mean \(\langle n\rangle_\mathrm{fit}\), obtained
from the Poisson description of the fitted component areas.
Device~A is shown on a linear scale,
whereas Device~B is shown on a logarithmic scale to display its broader
dynamic range. Colored curves show the individual EMG components,
labeled by \(n\), and the red curves show the total fits, including a 
constant background term. Peaks are resolved through
\(n=5\) for Device~A and through \(n=8\) for Device~B.}
\label{fig:2}
\end{figure*}

To parameterize this edge-distance dependence, we introduce the minimal
phenomenological barrier form
\begin{equation}
U_\mathrm{nuc}(\tilde{x})
=
c_\mathrm{nuc}\,\tilde{x}^{p_\mathrm{nuc}}\,\kb T,
\label{eq:barrier_nuc}
\end{equation}
so that the corresponding waiting-time scale is
\begin{equation}
T_\mathrm{nuc}(\tilde{x})
=
\tau_0\exp\!\left[
c_\mathrm{nuc}\tilde{x}^{p_\mathrm{nuc}}
\right].
\label{eq:Tnuc}
\end{equation}
Here \(c_\mathrm{nuc}\) and \(p_\mathrm{nuc}\) are fit parameters. We
relate the EMG exponential-tail scale to the mean
nucleation-associated waiting-time scale,
\(\tau_n=\mathbb{E}[T_\mathrm{nuc}(\xmin)]\), over the nearest-edge
distribution:
\begin{empheq}[box=\fbox]{equation}
\begin{aligned}
\tau_n
=
\tau_0\,n\int_0^1
(1-\tilde{x})^{n-1}
\exp\!\left[
c_\mathrm{nuc}\tilde{x}^{p_\mathrm{nuc}}
\right]\,d\tilde{x}.
\end{aligned}
\label{eq:tau_exact}
\end{empheq}
This phenomenological expression parameterizes the measured decrease of
\(\tau_n\) with photon number.

For the traversal contribution, let \(d_\mathrm{eff}\) denote the
effective transverse redistribution length associated with each absorbed
photon, and define
\begin{equation}
\beta=\frac{d_\mathrm{eff}}{W}.
\label{eq:beta}
\end{equation}
In this same approximation, each absorbed photon is represented by an
effective blocked interval of transverse width \(d_\mathrm{eff}\). In
the one-dimensional collective limit used here, the absorption-induced
perturbations are taken to act at the same position along the strip
while occupying non-overlapping transverse intervals. Current
conservation then gives the effective relation
\begin{equation}
j_\mathrm{gap}(n)
=
\frac{j_0}{1-n\beta},
\qquad
\alpha\equiv\frac{j_0}{j_c},
\label{eq:jgap}
\end{equation}
where \(j_0\) is the nominal bias-current density and \(j_c\) is the
characteristic current density at which the traversal barrier vanishes
in this parameterization. This approximation is used only in the regime
\(n\beta<1\).

We parameterize the effective traversal barrier as
\begin{equation}
U_\mathrm{trav}(n)
=
U_0\left(
1-\frac{\alpha}{1-n\beta}
\right),
\label{eq:barrier_trav}
\end{equation}
where \(U_0\) is an effective zero-current barrier scale. Using an
activated dependence and absorbing photon-number-independent factors
into the prefactor gives
\begin{empheq}[box=\fbox]{equation}
\mu_n
=
\mu_c+
T_0\exp\!\left[
-\frac{\gamma n}{1-n\beta}
\right],
\qquad
\gamma=
\alpha\beta\frac{U_0}{\kb T}.
\label{eq:mu_exact}
\end{empheq}
Here \(T_0\) and \(\gamma\) are fit parameters, whereas
\(\alpha=j_0/j_c\) is fixed by the bias condition.
For the timing fits, \(d_\mathrm{eff}\) is held fixed at
\SI{25}{nm}, motivated by the hotspot-interaction scale reported in
Ref.~\cite{renema2017}, with \(\beta=d_\mathrm{eff}/W\).
This adopted length is not determined by the timing fits and should
not be interpreted as a measurement of the collective redistribution
range. Thus, \(\mu_n-\mu_c\) represents an effective traversal
contribution rather than a directly measured vortex-crossing time.

The remaining EMG parameter, \(\sigma_n\), describes the symmetric
Gaussian broadening of each TOA component. It includes contributions
from the finite laser pulse width, synchronization and TCSPC timing,
electrical readout noise, and intrinsic detector timing fluctuations.
We therefore treat \(\sigma_n\) empirically and do not assign it to
either the nucleation or traversal contribution.

Together, \(\mu_n\), \(\sigma_n\), and \(\tau_n\) determine the shapes
and overlap of adjacent fitted EMG components. The separation observables
used to quantify adjacent-photon-number TOA discrimination are defined in
Sec.~\ref{sec:results}.

\section{Results}
\label{sec:results}

This section presents the measured TOA histograms and the timing
parameters extracted from the EMG decomposition. We first establish the
component assignments and the identifiable sequence in each device. We
then examine the photon-number dependence of the exponential-tail scale
\(\tau_n\), the Gaussian-location parameter \(\mu_n\), and the Gaussian
width \(\sigma_n\), before combining the fitted peak times and widths to
quantify adjacent-component discrimination.

We define the FWHM-normalized separation of adjacent fitted TOA
components as
\begin{equation}
\mathcal{R}_{n,n+1}^{\mathrm{FWHM}}=
\frac{\left|t_{n+1}^{\mathrm{pk}}-t_n^{\mathrm{pk}}\right|}
{\left[\mathrm{FWHM}_n+\mathrm{FWHM}_{n+1}\right]/2},
\label{eq:fwhm_separation}
\end{equation}
where the numerator is the temporal separation between the two fitted
peak times and the denominator is the mean FWHM of the corresponding
components~\cite{schapeler2025}. We use
\(\mathcal{R}_{n,n+1}^{\mathrm{FWHM}}>1\) as a descriptive
FWHM-separation benchmark.

We further quantify adjacent-component separation through the
equal-prior pairwise TOA-assignment fidelity,
\begin{equation}
F_{n,n+1}^{\mathrm{pair}}=
1-\frac{1}{2}
\int_{-\infty}^{\infty}
\min\!\left[p_n(t),p_{n+1}(t)\right]dt.
\label{eq:pairwise_fidelity}
\end{equation}
The integral is the overlap between the two normalized fitted TOA
distributions. For equal prior probabilities,
\(F_{n,n+1}^{\mathrm{pair}}\) is the optimal single-event assignment
fidelity obtainable from the fitted TOA distributions of the specified
adjacent components~\cite{schapeler2026}.

Representative TOA histograms for Devices~A and B are shown in
Fig.~\ref{fig:2}. In both devices, the EMG mixture of
Eq.~\eqref{eq:hist_model} describes the measured distributions and
yields \((\mu_n,\sigma_n,\tau_n)\) for each fitted component. 
As the incident optical power increases, the relative populations of the
components shift toward higher photon numbers. Their normalized areas
follow the conditional Poisson description expected for coherent-state
illumination, supporting the assignment of the resolved timing components
to photon-number classes. The fitted Poisson mean,
\(\langle n\rangle_\mathrm{fit}\), increases with optical power. The
Poisson fits are provided in the Supplementary Material.

For a given photon-number component, the fitted timing parameters remain
broadly stable over the optical-power range in which the corresponding
component is well populated. Increasing the incident optical power
primarily redistributes the component populations, while the fitted
parameters retain their characteristic photon-number dependence.

Both devices retain a clear component assigned to \(n=1\). The finite
sequence of resolved peaks extends through \(n=5\) in Device~A and
through \(n=8\) in Device~B. Increasing the incident optical power
changes the populations of these components without revealing
additional peaks. Under the present illumination, bias, and
fitting conditions, the wider device therefore exhibits a longer
resolved TOA-peak sequence.

\begin{figure}[h]
\centering
\includegraphics[width=0.98\columnwidth]{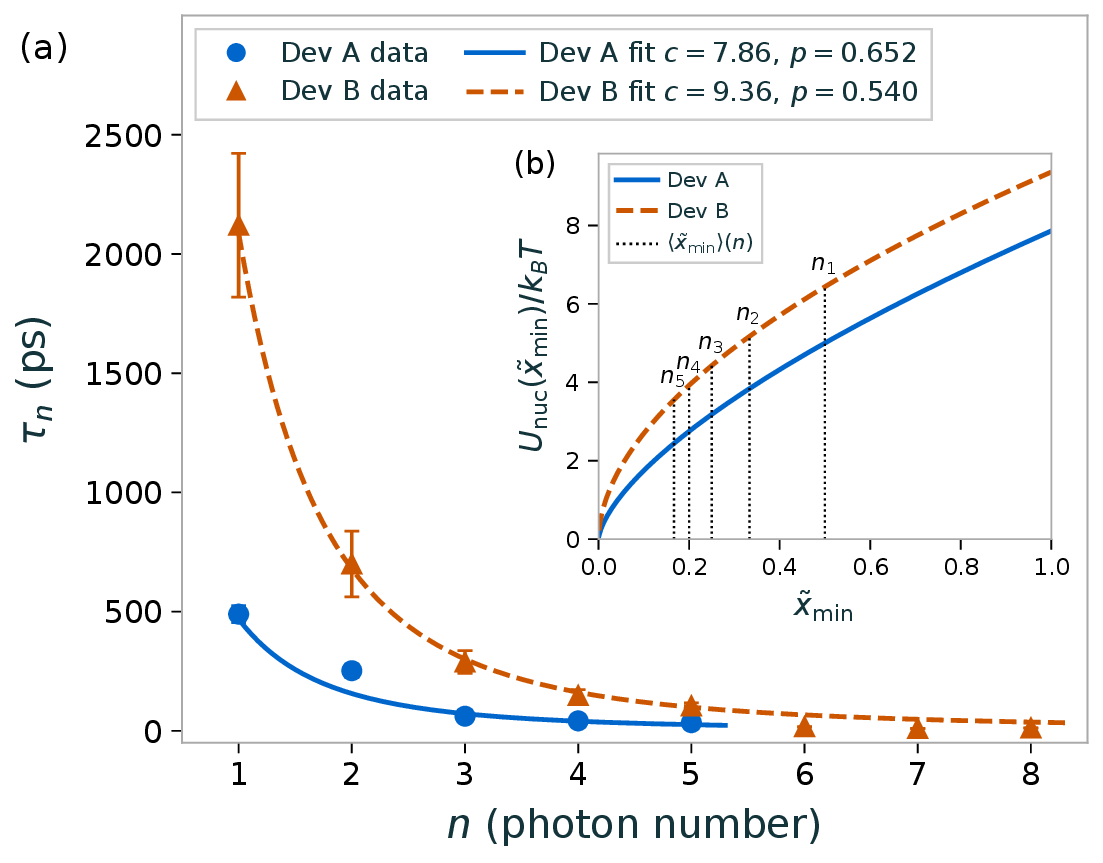}
\caption{Photon-number dependence of the phenomenological
edge-sensitive contribution.
(a) Extracted exponential-tail scale \(\tau_n\) for Device~A (blue circles) and
Device~B (orange triangles), together with fits to Eq.~\eqref{eq:tau_exact}.
(b) Reconstructed nucleation-barrier proxies
\(U_\mathrm{nuc}/k_B T=c_\mathrm{nuc}\tilde{x}^{p_\mathrm{nuc}}\) using the
best-fit values of \(c_\mathrm{nuc}\) and \(p_\mathrm{nuc}\). The markers
indicate the expected mean position of the edge-nearest hotspot,
\(\langle \xmin \rangle = 1/(n+1)\), for the resolved photon numbers.}
\label{fig:3}
\end{figure}

\begin{figure}[h]
\centering
\includegraphics[width=0.98\columnwidth]{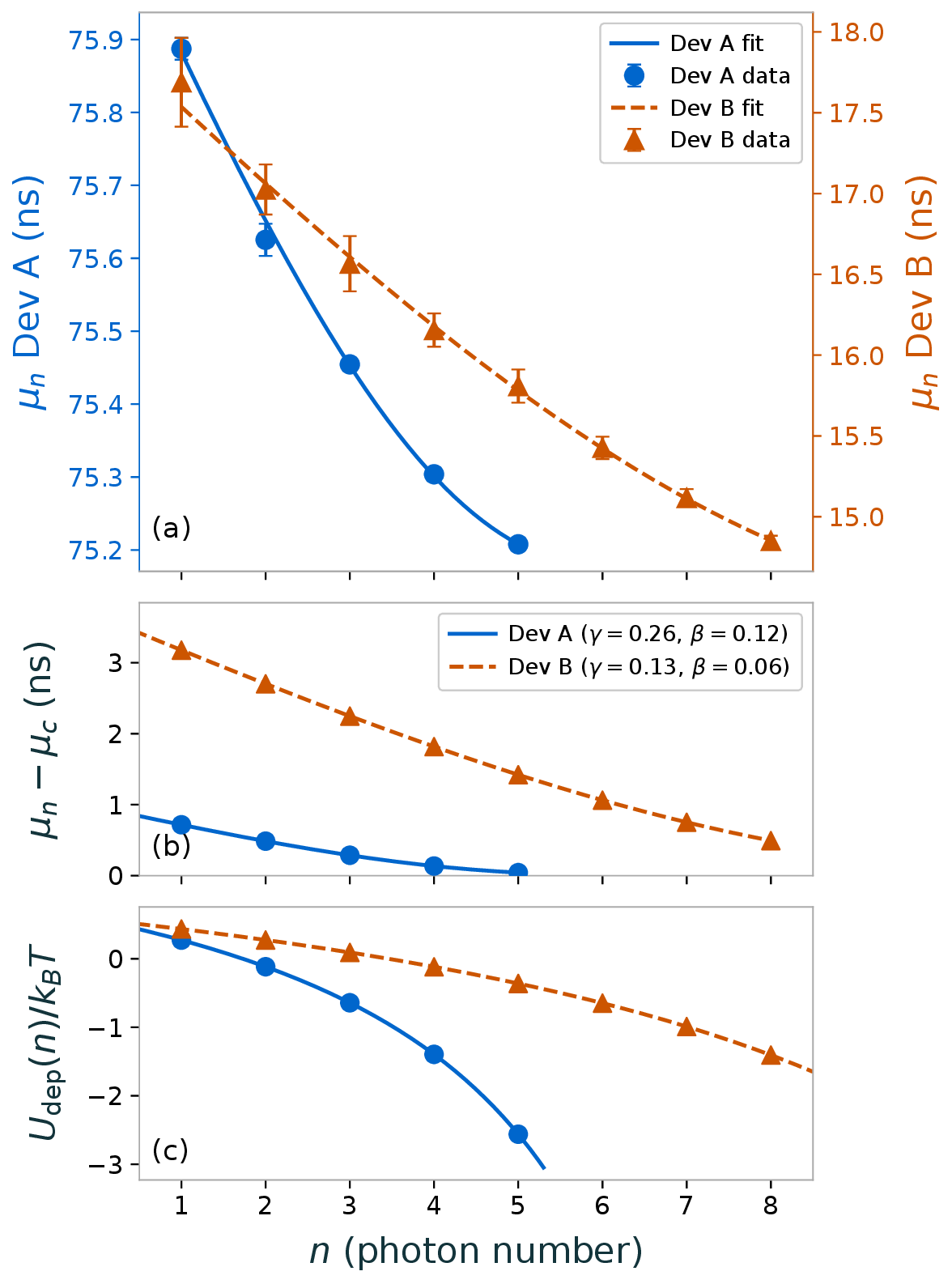}
\caption{Photon-number dependence of the phenomenological collective
redistribution contribution.
(a) Extracted Gaussian-location parameter \(\mu_n\) for Device~A
(blue circles, solid) and Device~B (orange triangles, dashed), together
with fits to Eq.~\eqref{eq:mu_exact}. (b) Effective traversal contribution
obtained from the fitted model after removal of the device-specific timing offset.
(c) Reconstructed traversal-barrier proxy
\(U_\mathrm{trav}(n)/k_B T\) as a function of photon number.}
\label{fig:4}
\end{figure}

\begin{figure}[t]
\centering
\includegraphics[width=\linewidth]{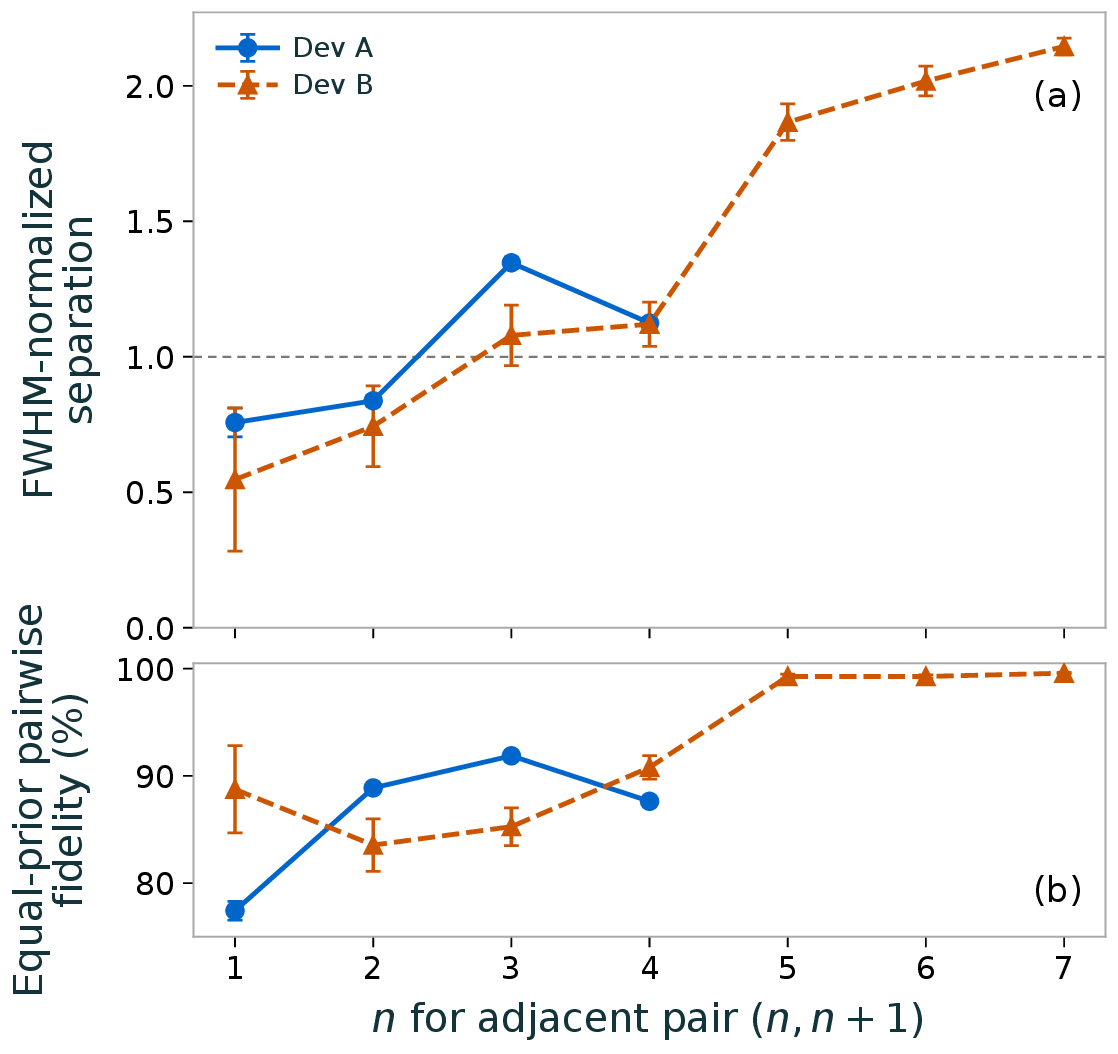}
\caption{Adjacent-component TOA discrimination in Devices~A and B.
(a) FWHM-normalized peak separation
\(\mathcal{R}_{n,n+1}^{\mathrm{FWHM}}\).
(b) Equal-prior pairwise TOA-assignment fidelity
\(F_{n,n+1}^{\mathrm{pair}}\). Here \(n\) denotes the lower photon number 
in the pair \((n,n+1)\). For Device~B, the components assigned to
\(n=7\) and \(n=8\) give
\(\mathcal{R}_{7,8}^{\mathrm{FWHM}}=2.15\pm0.03\) and
\(F_{7,8}^{\mathrm{pair}}=(99.58\pm0.06)\%\). Lines are guides to the
eye.}
\label{fig:5}
\end{figure}

We next examine the photon-number dependence of the fitted timing
parameters and adjacent-component discrimination. The experimental
values shown in Figs.~\ref{fig:3}--\ref{fig:5} are averages over the
measured optical-power range.
The exponential-tail scale \(\tau_n\), shown in
Fig.~\ref{fig:3}(a), decreases monotonically with photon number
in both devices.
For Device~A, it falls from
\(\tau_1\approx\SI{490}{ps}\) to
\(\tau_5\approx\SI{34}{ps}\), a reduction by approximately a factor of
14. Device~B starts from a larger value,
\(\tau_1\approx\SI{2120}{ps}\), and decreases over the components
assigned through \(n=8\). Equation~\eqref{eq:tau_exact} reproduces
these trends with \(c_\mathrm{nuc}=7.86\) and
\(p_\mathrm{nuc}=0.65\) for Device~A, and
\(c_\mathrm{nuc}=9.36\) and \(p_\mathrm{nuc}=0.54\) for Device~B.

The Gaussian-location parameter \(\mu_n\), shown in
Fig.~\ref{fig:4}(a), decreases monotonically with photon number
in both devices, but more gradually than \(\tau_n\).
Device~A shows progressively compressed \(\mu_n\) spacing at larger
\(n\), whereas Device~B retains a comparatively uniform sequence
through the component assigned to \(n=8\). Equation~\eqref{eq:mu_exact}
describes the measured \(\mu_n\) values with
\(T_0=\SI{966}{ps}\) and \(\gamma=0.26\) for Device~A, and
\(T_0=\SI{3664}{ps}\) and \(\gamma=0.14\) for Device~B.
The different fitted offsets are expected from the cable and readout
delays used in the two measurements. Subtracting these offsets from
the fitted curves gives the effective traversal contributions shown in
Fig.~\ref{fig:4}(b).

To visualize the phenomenological interpretation of these timing trends,
Figs.~\ref{fig:3}(b) and \ref{fig:4}(c) show the corresponding
model-derived barrier proxies,
\(U_\mathrm{nuc}(\tilde{x})/k_B T\) and
\(U_\mathrm{trav}(n)/k_B T\), respectively. 
These profiles summarize the parameterizations used to describe the
fitted \(\tau_n\) and \(\mu_n\) trends.

The remaining timing parameter is the Gaussian width \(\sigma_n\). In
both devices, \(\sigma_n\) decreases with photon number and approaches
a floor at larger \(n\). It includes contributions from the laser pulse
width, synchronization and TCSPC timing, electrical readout noise, and
intrinsic detector timing fluctuations. The full \(\sigma_n\) data and
fit uncertainties are provided in the Supplementary Material.

The combined evolution of \(\tau_n\), \(\mu_n\), and \(\sigma_n\)
determines the shapes and overlap of adjacent fitted TOA components.
Figure~\ref{fig:5} shows that the FWHM-normalized
separation \(\mathcal{R}_{n,n+1}^{\mathrm{FWHM}}\) generally increases
with photon number. For both devices, adjacent pairs with \(n\geq3\) satisfy
\(\mathcal{R}_{n,n+1}^{\mathrm{FWHM}}>1\). The largest measured
separation occurs for the components assigned to \(n=7\) and \(n=8\) in
Device~B, with
\(\mathcal{R}_{7,8}^{\mathrm{FWHM}}=2.15\pm0.03\) and an equal-prior
pairwise TOA-assignment fidelity of
\(F_{7,8}^{\mathrm{pair}}=(99.58\pm0.06)\%\).

This behavior contrasts with the compression of neighboring timing
shifts at high photon number reported for narrow-wire SNSPDs
~\cite{sidorova2025}. In those devices, a characteristic timing
contribution scales approximately as \(n^{-1/2}\), so the spacing
between neighboring Gaussian-location parameters decreases as
\(\Delta\mu_n\equiv\mu_n-\mu_{n+1}\propto
n^{-1/2}-(n+1)^{-1/2}\). In the present wide SMSPDs, neighboring TOA
peaks remain distinguishable at high photon number, with their
separation relative to their widths generally increasing across the
resolved range.

\section{Discussion}
\label{sec:discussion}

The central observation is that neighboring TOA peaks become more
distinguishable relative to their widths at higher photon number.
The EMG tail scale \(\tau_n\) decreases strongly across the resolved
range, while the Gaussian locations \(\mu_n\) remain distinct for
neighboring components, particularly in Device~B. Thus, the improved
FWHM-normalized separation reflects the combined evolution of peak
positions and component widths. 
In the phenomenological model, the decrease of \(\tau_n\) reflects
the increasing likelihood of photon absorption near a strip edge,
whereas the continued separation of neighboring \(\mu_n\) values is
associated with current redistribution caused by multiple
photon-induced hotspots.

Within the phenomenological description, collective current
redistribution provides a possible explanation for the finite sequence
of resolved TOA peaks. As the absorption-induced perturbations occupy
a larger fraction of the strip width, the effective current density
in the remaining superconducting region increases. The associated
suppression of the traversal barrier can lead to saturation of the
timing response, beyond which additional absorbed photons no longer
produce distinguishable TOA components. The longer resolved sequence
in the wider device is qualitatively consistent with this picture.

Separately, the observed termination can be used to estimate an
effective collective redistribution scale. This estimate is distinct
from the fixed \SI{25}{nm} length adopted in the timing fits.
If the highest resolved photon number \(N_\mathrm{max}\) marks an
effective perturbed fraction of order unity, then
\[
d_\mathrm{eff}\sim\frac{W}{N_\mathrm{max}}.
\]
For the present devices, this gives approximately
\(0.6\text{--}0.8~\si{\um}\). This is a model-dependent estimate of
the effective collective redistribution scale, conditional on
attributing the observed termination to this mechanism.

\begin{figure}[h]
\centering
\includegraphics[width=\linewidth]{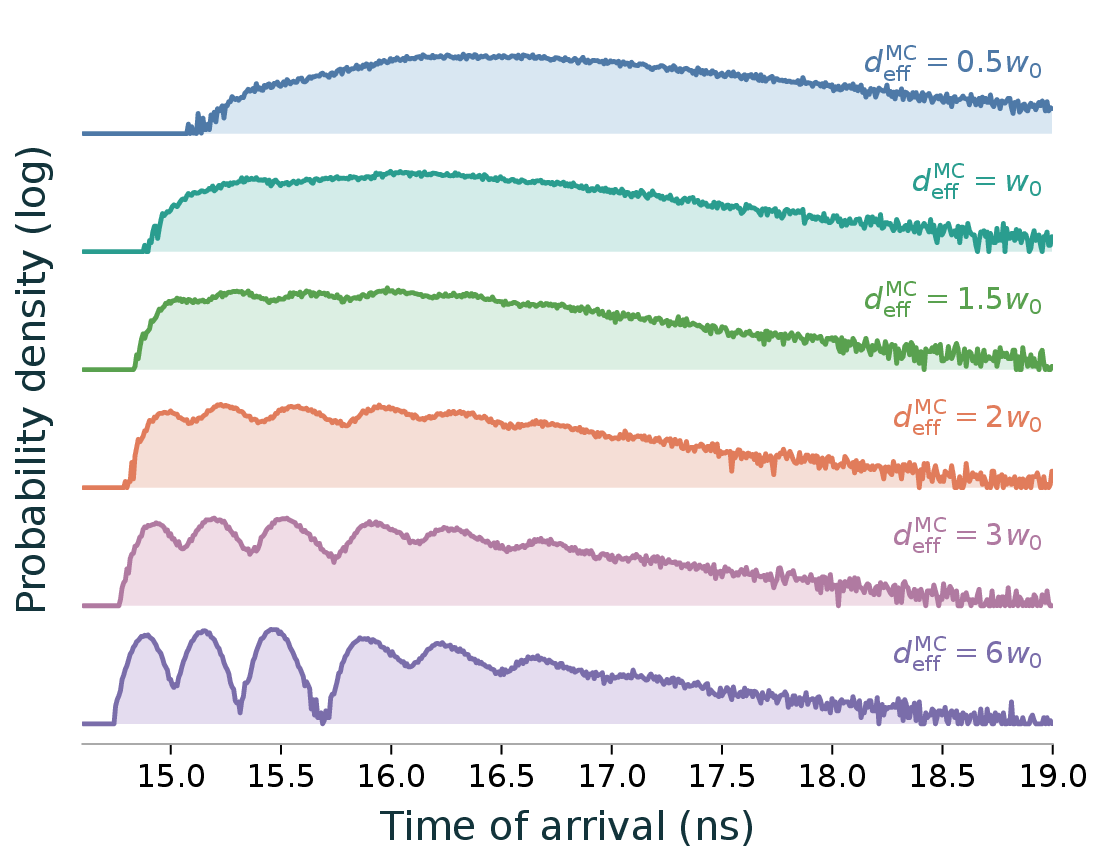}
\caption{Monte Carlo TOA distributions for Device~B on a log scale for
different effective redistribution ranges, denoted as
\(d_\mathrm{eff}^{\mathrm{MC}}\) in the text, relative to the
optical beam waist \(w_0\). As
\(d_\mathrm{eff}^{\mathrm{MC}}/w_0\) increases, the timing response
approaches the one-dimensional collective limit and the discrete
photon-number structure becomes more clearly resolved. For smaller
\(d_\mathrm{eff}^{\mathrm{MC}}/w_0\), the valleys between adjacent
peaks become less visible.}
\label{fig:6}
\end{figure}

We further consider the effect of finite photon-beam width. 
The preceding description neglects the longitudinal separation of
absorption sites, whereas the experiment uses a focused optical beam
with a two-dimensional spatial profile and waist \(w_0\).
In the Monte Carlo model, absorption positions are sampled from the beam 
profile in both transverse and longitudinal directions, and each absorption 
event is assigned an effective redistribution range
\(d_\mathrm{eff}^{\mathrm{MC}}\). The collective response depends on
how often multiple perturbations overlap within the illuminated area of the device.
When their longitudinal positions are sufficiently close, the response
approaches the one-dimensional collective limit and the multi-peak TOA
structure becomes more clearly resolved. As shown in
Fig.~\ref{fig:6}, the simulated timing structure is strongly
smeared for \(d_\mathrm{eff}^{\mathrm{MC}}\lesssim w_0\), where
better peak separation emerges for approximately
\(d_\mathrm{eff}^{\mathrm{MC}}\gtrsim2w_0\). With
\(w_0\approx\SI{0.3}{\um}\), this suggests a lower effective scale of
\(d_\mathrm{eff}^{\mathrm{MC}}\gtrsim\SI{0.6}{\um}\). 

The estimate of \(d_\mathrm{eff}\) inferred from the termination of the
resolved TOA-peak sequence and the value of
\(d_\mathrm{eff}^{\mathrm{MC}}\) required in the Monte Carlo
calculation are obtained by different treatments of the same
phenomenological picture. Both place the effective
scale relevant to the present timing picture at several hundred
nanometers, larger than the hotspot interaction length of order
\SI{25}{nm} reported for NbN nanowire devices
~\cite{renema2017}.

Our work is intended to describe the measured timing trends
rather than identify the microscopic origin of each fitted
parameter. Hotspot formation and relaxation, quasiparticle diffusion,
vortex dynamics, and the quantum nature of the detection process may
all contribute to the observed TOA response. Spatially scanned
illumination, controlled beam-waist measurements, bias-dependent and
wavelength-dependent timing measurements, independently calibrated
photon occupancy, and timing-chain characterization would help clarify
their roles in the TOA response and in the termination of the resolved
peak sequence.

From an application perspective, wide SMSPDs are promising for intrinsic
single-element TOA-based photon-number discrimination because the
adjacent-component separation improves over the resolved range. Reducing
latching could allow operation closer to the switching current and
improve the low-photon-number response, where the large \(\tau_n\) contribution
still limits discrimination. Edge engineering may promote more
reproducible early detection dynamics and suppress the broad low-photon-number
tail. At higher photon number, width and optical-mode optimization may
help preserve the evolution of \(\mu_n\) across a broader range. Thus, 
wide SMSPDs offer both a candidate route toward intrinsic
single-element photon-number discrimination and a useful platform for
studying the interplay between local photon absorption, hotspot
formation and evolution, and vortex-assisted detection.

\section{Conclusion}

We have measured photon-number-dependent TOA distributions in wide,
short NbN SMSPDs under pulsed \SI{515}{nm} excitation. The
\SI{3.2}{\um}- and \SI{6.4}{\um}-wide devices exhibit resolved EMG timing
components assigned through \(n=5\) and \(n=8\), respectively. In both
devices, the FWHM-normalized separation between adjacent TOA peaks
generally increases with photon number. For the components assigned to
\(n=7\) and \(n=8\) in the wider device, this separation reaches
\(\mathcal{R}_{7,8}^{\mathrm{FWHM}}=2.15\pm0.03\), and the fitted EMG
distributions yield an equal-prior pairwise TOA-assignment fidelity of
\(F_{7,8}^{\mathrm{pair}}=(99.58\pm0.06)\%\).

With increasing photon number, the fitted TOA components narrow,
their exponential tails shorten, and their locations continue to shift.
A phenomenological model interprets the tail reduction and location
shifts in terms of edge-sensitive nucleation and collective current
redistribution, respectively.

Both devices also exhibit a finite sequence of distinguishable
components, even at the highest explored incident powers. Collective
current redistribution provides a possible explanation for this
termination, but the microscopic processes underlying the overall
timing response remain to be established. Further measurements with
controlled illumination and operating conditions, together with
independent photon-occupancy and timing-chain characterization, could
clarify these processes and the limits of photon-number discrimination.

These results support wide, short SMSPDs as a route toward intrinsic
single-element photon-number resolution. The improved relative TOA
separation at higher photon number motivates further device and readout
optimization for multiphoton-state characterization, while this geometry
provides a setting for investigating how the spatial distribution of
photon absorption influences superconducting detection dynamics.

\bibliography{transverse_smspd}

\end{document}